\documentstyle[prb,%preprint,
aps,psfig]{revtex}
\begin{document}
\draft

%\onecolumn

\title{Theoretical study of the chlorine adsorption on the Ag(111) surface}
\author{K. Doll$^{1,2}$ and N. M. Harrison$^{1,3}$}
\address{$^1$CLRC, Daresbury Laboratory, Daresbury, Warrington, WA4 4AD, UK}
\address{$^2$ Institut f\"ur Mathematische Physik, TU Braunschweig,
Mendelssohnstra{\ss}e 3, D-38106 Braunschweig}
\address{$^3$  Department of Chemistry, Imperial College 
of Science, Technology and Medicine, South Kensington, London, SW7 2AY, UK}
\maketitle

\begin{abstract}
We study the adsorption of chlorine on the Ag(111) surface with
full potential gradient corrected density functional calculations. 
When considering a $\sqrt{3} \times \sqrt{3}\ {\rm R}30^\circ$ pattern,
we find that the fcc hollow is the most favorable adsorption site.
We obtain an
Ag-Cl bond length of 2.62 \AA \ which is intermediate between
 two controversial
experimental results. We discuss the differences of core level energies and
densities of states for
the different adsorption sites.
We find the Cl-Ag interaction to be more consistent with an ionic, 
rather than
covalent, picture of the bonding.
In addition, we compute energetics and related
properties of Ag bulk and the clean Ag surface comparing the
local density approximation and the generalized gradient approximation.

\end{abstract}

\pacs{ }

\narrowtext
\section{Introduction}

The study of halogenide adsorption on metal surfaces has become a very
interesting field in surface science. Important questions such
as adsorption geometries and energetics have been addressed in 
experimental studies. The adsorption of chlorine on the Ag(111) surface
has been studied by various groups and with different techniques including
low energy electron diffraction (LEED) 
\cite{GoddardLambert,RovidaPratesi,BowkerWaugh,Eltsov,Shard}, 
surface extended X-ray 
absorption fine structure (SEXAFS) \cite{Lamble,Shard}, Auger spectroscopy
\cite{GoddardLambert,RovidaPratesi,BowkerWaugh}, 
thermal desorption\cite{GoddardLambert,BowkerWaugh},  
measurements of  adsorption isobars and isotherms \cite{TuBlakely},
work function measurements \cite{GoddardLambert} and
scanning tunneling microscopy \cite{Eltsov,SchottWhite}.

However, there is still an ongoing discussion on the correct 
interpretation of the experiments. Various explanations of observed
adsorption patterns have been proposed, and the
bond length has been measured twice with significantly different
results \cite{Lamble,Shard}.

We recently presented a theoretical study of the chlorine adsorption
on the Cu(111) surface \cite{DollHarrison2000} and
we are therefore confident that simulation can shed some light on
the difficult question of chlorine
adsorption. As a test of our computational method, we performed additionally
calculations on Ag bulk and the clean surfaces. These are presented in
section  \ref{Agbulksection}. The question of chlorine adsorption is discussed
 in section \ref{chlorineadsorptionsection}.

\section{Ag bulk and clean surfaces}
\label{Agbulksection}

\subsection{Computational Parameters}

We used a local basis set formalism where the basis functions are
Gaussian type orbitals centered at the atoms as implemented in the code
 CRYSTAL \cite{Manual}. In order to examine the effect of various 
treatments of electron exchange and correlation, we have
used  the local density approximation
(LDA) with Dirac-Slater exchange \cite{LDA} 
and the Perdew-Zunger correlation functional \cite{PerdewZunger}, 
the gradient corrected exchange and correlation functional of Perdew and Wang
(GGA-PW) \cite{Perdewetal} and the gradient corrected functional of
Perdew, Burke and Ernzerhof (GGA-PBE) \cite{PBE}.

For silver, we used a relativistic pseudopotential with corresponding 
basis set \cite{Andrae}. For the calculation on the free atom, we used
the original basis set. For the calculation on the solid and surfaces,
we kept the inner  exponents ($[2s1p1d]$) fixed and re-optimized the outermost
exponents within the GGA-PW 
approximation. Values of 0.63 $\frac{1}{a_0^2}$ (with the Bohr
radius $a_0$) and 0.11
$\frac{1}{a_0^2}$ for the
$s$-exponents, 0.23 $\frac{1}{a_0^2}$ for the $p$-exponent and 
0.18 $\frac{1}{a_0^2}$ for the $d$-exponent
were obtained (the most diffuse $s$-exponent could not be optimized and
was kept fixed at a value of 0.11 $\frac{1}{a_0^2}$ 
to avoid numerical instabilities).
This resulted as a whole in a $[4s3p2d]$ basis set.
The Cl basis set is $[5s4p1d]$ as used in our recent study of Cl adsorption
 on Cu(111)
\cite{DollHarrison2000}.

The density functional potential was fitted with an auxiliary basis set,
which consisted, for both Ag and Cl, of 12  $s$ and $p$ functions with
exponents taken to be a geometrical sequence from  0.1 $\frac{1}{a_0^2}$
 to 2000 $\frac{1}{a_0^2}$. For Cl
5 $d$ functions with exponents in the range 0.8 $\frac{1}{a_0^2}$ 
to 100 $\frac{1}{a_0^2}$ were employed,
for Ag 8 $d$ and $f$ functions from 0.1 $\frac{1}{a_0^2}$ to 100 
$\frac{1}{a_0^2}$ . 

We used a Pack-Monkhorst net of
shrinking factor 16 for Brillouin zone sampling, which
resulted in 145 irreducible $\vec k$-points for the bulk, 
45 for the clean (100) surface,
81 for the clean (110) surface,
30 for the clean (111) surface, 
and 30 or 73 for the system Cl/Ag(111), depending on the
adsorption site (30 for hcp, 73 for fcc, bridge and atop site --- an
explanation of the sites is given in section \ref{chlorineadsorptionsection};
see also figure \ref{fort34-figure} \cite{DLV}). A smearing
temperature of 0.03 $E_h$ was applied to the Fermi function ($E_h$=27.2114 eV).
This approach has been shown to give stable results for lithium bulk
and surfaces
\cite{KlausNicVic} as well as for the study of Cl on the Cu(111) surface
\cite{DollHarrison2000}.

\subsection{Properties of bulk Ag and clean Ag surfaces}

Results for cohesive properties of bulk Ag are  in excellent agreement with 
those published in the literature (table \ref{Agbulktable}).
The tendency of the LDA to overbind results in a short lattice constant, 
too high
bulk modulus and too high binding energy. GGA improves the binding
energy and bulk modulus; the deviation of the lattice constant is similar
to LDA (with opposite sign, however).

Surface formation energies (in $E_h$ per surface atom)
of the clean, unrelaxed surfaces are displayed in table
\ref{Agsurfsummtable}. The surfaces were modeled with a finite number of
Ag layers (typically from 1 to 7); 
this model was $not$ periodically repeated in space. 
The surface energy was found to be stable from three layers onwards.
The bulk energy was computed with the same computational parameters
(basis set, shrinking factors, smearing temperature).

LDA binding energies are approximatively 50\%
higher than those obtained with gradient corrected functionals, which is
higher than found in a previous study \cite{Vitosetal}. 
GGA-PW and GGA-PBE turned out to give practically identical results in
all cases which are in excellent agreement with the 
literature \cite{Vitosetal}.

\section{Cl on the Ag(111) surface}
\label{chlorineadsorptionsection}

The adsorption of Cl on the Ag(111) surface has been investigated by
several groups. Still, there is no agreement on adsorption patterns,
coverages, 
preferred sites and bond lengths. In the first study with LEED and
Auger spectroscopy \cite{RovidaPratesi}, a 
$\sqrt{3} \times \sqrt{3}\ {\rm R}30^\circ$ pattern at low chlorine
coverage and a $(3 \ \times 3)$ pattern at higher coverage were observed.
In a second study, a $\sqrt{3} \times \sqrt{3}\ {\rm R}30^\circ$ pattern
at a coverage of 1/3 
and a $(10 \ \times 10)$ pattern at a coverage of 0.49
were found \cite{GoddardLambert}.
Within an Arrhenius model, the
 binding energy was determined to be 0.080 $E_h/{\rm Cl_2}$ .
A similar energy was found in another experiment\cite{TuBlakely}, but 
the LEED pattern was interpreted as an epitaxial AgCl layer. 
A $\sqrt{3} \times \sqrt{3}\ {\rm R}30^\circ$ pattern at a coverage close to
one was found in ref. \onlinecite{BowkerWaugh}. 
Scanning tunneling microscopy
 experiments have been interpreted as a parallel double row structure,
where Cl atoms occupy bridge and threefold hollow sites \cite{SchottWhite},
or as $(17 \times 17)$ overlayer at saturation \cite{Eltsov}.

A structural determination with SEXAFS gave a Ag-Cl 
bond length of 2.70 $\pm$ 0.02 \AA \ at a nominal coverage of 1/3 of a
monolayer and 
2.70 $\pm$ 0.01 \AA \ at a nominal coverage of 2/3 of a 
monolayer\cite{Lamble}.
Cl was found to adsorb in threefold hollow sites in a 
$\sqrt{3} \times \sqrt{3}\ {\rm R}30^\circ$ pattern, for the higher coverage
a vacancy honeycomb structure was proposed.
This determination has been
 questioned in a recent article where a bond length of 
2.48$\pm$0.04 \AA \ has been deduced from
 SEXAFS measurements for a nominal coverage of 1/3 of a monolayer\cite{Shard}. 
In this study an order-disorder
transition was found at 195 K, below which a
 $\sqrt{3} \times \sqrt{3}\ {\rm R}30^\circ$ pattern at a coverage
of 1/3 of a monolayer was identified in LEED measurements; this pattern
was found to vanish above the transition temperature. 
In addition, a $(13 \times 13)$ overlayer was identified at a coverage
of 0.37. 
The bond length of 2.48 \AA \ was found to fit better with other bond lengths
of chlorine adsorbates on 3$d$ and 4$d$ fcc metals. Therefore, 
it was suggested that previous experiments should be reinterpreted.
A variation of the bond length with coverage was neither
 found for Cl on Ag(111)
\cite{Lamble}
nor for the similar system of I on Rh(111) \cite{Barnes}.

\begin{figure}
\caption{The structures considered for Cl, 
adsorbed on the Ag(111) surface at one
third coverage, in a $\sqrt{3} \times \sqrt{3}\ {\rm R}30^\circ$ unit cell.
When Cl is adsorbed in an fcc hollow, it sits above a Ag atom in the third
layer (upper left) while in the hcp hollow it is above an atom in the 
second layer (upper right).
The atop position is Cl adsorbed vertically above a surface atom
(lower left).
In the bridge position it is vertically above the middle of two
surface atoms (lower right). The $x$-direction is horizontal in the
plane, the $y$-direction is vertical, and the $z$ direction points out of
the plane of the page.}
\label{fort34-figure}

\centerline
{\psfig
{figure=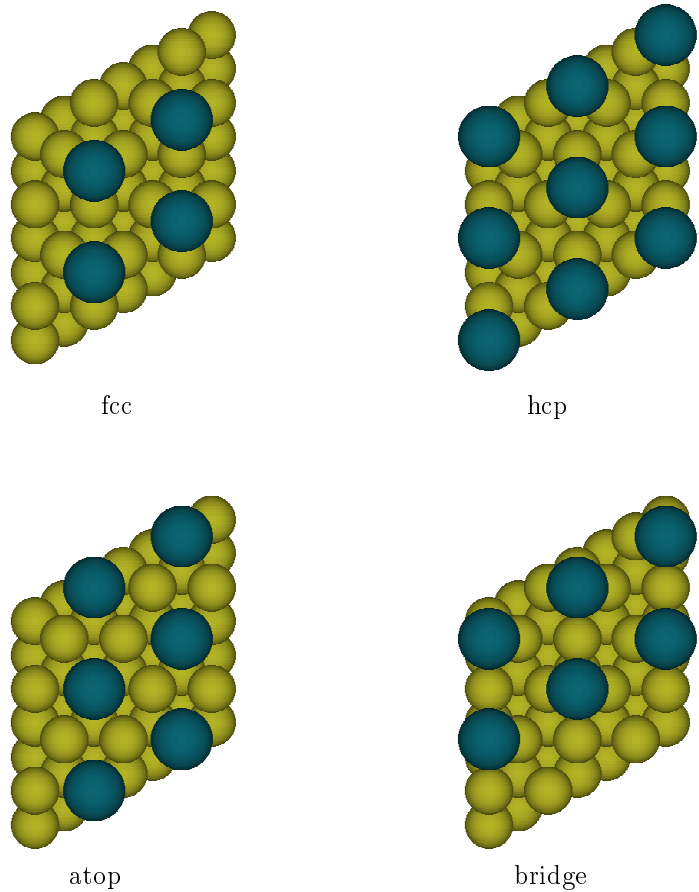,width=15cm,angle=0}}
\end{figure}

These experimental observations therefore pose an interesting problem for 
simulation studies. In our calculations,
we modeled the adsorption of Cl on Ag(111) as a slab periodic in two
dimensions but with a finite thickness
of three or four layers of silver.
This was found
to be sufficiently thick to obtain stable results for the surface energy
of the clean Ag(111) surface.
The slab was covered by
a layer of chlorine corresponding to a coverage of one third of a 
monolayer, in a $\sqrt{3} \times \sqrt{3}\ {\rm R}30^\circ$ pattern.
Cl was adsorbed in four different sites: the atop site on top of a silver atom
in the top layer, the bridge site in the middle of two Ag in the top layer,
and hcp (and fcc) hollows where the chlorine sits in threefold hollow sites
with a silver atom in the second (third) layer under the chlorine atom (see
figure \ref{fort34-figure}). In the following, all the calculations were 
performed 
at the GGA-PW level, if not explicitly stated otherwise.

The results of these calculations are summarized in table
\ref{ClonAgtable}. First, it turns out that the threefold hollow sites
are lowest in energy, with the bridge site being about 0.003 $E_h$ higher
and the atop site 0.017 $E_h$ higher in energy. This picture is quite similar
to that of chlorine adsorption
 on the Cu(111) surface \cite{DollHarrison2000}. As the threefold sites are
nearly degenerate in energy, we considered an improved model with four
silver layers to investigate the dependence of the results with respect
to the number of layers. It turns out that the fcc hollow stays lower in
energy, although the energy difference is reduced from 0.001 $E_h$ (three
layers) to only 0.0004 $E_h$ (four layers). Therefore, a
fully reliable prediction of the adsorption site can not be made as more layers
might still change this picture. The total adsorption energy is 
0.111 $E_h$ per chlorine atom. We computed 
a dissociation energy $D_0^0=0.099$ $E_h$, vibrational constant $\omega_e=520 
\frac{1}{cm}$ and internuclear distance 
$r_e=2.05$ \AA \ for the chlorine dimer 
which is in reasonable agreement with experiment 
($D_0^0$=0.091 $E_h$, $\omega_e=560 \frac{1}{cm}$, $r_e$=1.99 \AA)
\cite{HuberHerzberg}.
Therefore, our binding energy corresponds
to 0.123 $E_h$ per $\rm Cl_2$
 when computed with respect to the $\rm Cl_2$ molecule,
using our computed binding energy for the chlorine dimer.
This is higher than that deduced from thermal desorption
experiments  \cite{GoddardLambert}(0.080 $E_h$)
 and measurements of adsorption isobars
and isotherms \cite{TuBlakely}
 ($\sim$ 0.088 $E_h$ for the heat of adsorption of
Cl$_2$ ). 
However,
the first experimental value depends on an empirical prefactor assumed in the
Arrhenius model and was questioned in ref. \onlinecite{GoddardLambert}; the
authors
stated that it should be higher as $D_0^0$ of the $\rm Cl_2$ molecule
because desorption as atomic Cl was observed.

Another important aspect is the adsorption geometry. We allowed the
top silver layer to relax vertically and a slight inwards relaxation by
about 0.04 \AA \ was found for all adsorption sites. The Ag-Cl bond length
was determined to be 2.62 \AA \ for the threefold hollow sites,
2.54 \AA \ for the bridge site and 2.38 \AA \ for the atop site.
Tests on the 4-layer slab demonstrated that these bond lengths are
stable with respect to the number of layers.
Other tests such as varying the outermost diffuse exponents in the basis set
or
applying a smaller smearing temperature were performed, and it turned
out that the bond length was also stable with respect to these parameters;
the variation was of the order $\pm$ 0.02 \AA.
Therefore, we can conclude that the computed bond length is not highly
sensitive to computational details. The remaining
 approximation in our simulations
is the choice of the
exchange-correlation functional. The effects of this choice were
examined by recomputing the surface properties within the LDA
 (table \ref{ClonAgtableLDA}).
LDA gives the same preference of adsorption sites and a similar energy
splitting.
The adsorption energies are 20 - 25 $mE_h$
higher than those computed using the GGA-PW, and the bond lengths are
shorter by 0.05 - 0.07 \AA.
The bond length computed with our model (2.62 \AA \ in GGA-PW, 2.55 \AA \
 in LDA) 
is intermediate between two
controversial experiments (2.48 \AA \cite{Shard} and 
2.70 \AA \cite{Lamble}); therefore we cannot support either
 experiment over the other.

Combining our data for Cl on Ag(111) and Cl on Cu(111), we can comment
on the effective radius of the Cl adsorbate, under the assumption that
both the chlorine adsorbate and the copper and silver surfaces are made
of hard spheres.
For the fcc site and Cl on Cu, a bond length of 2.40 \AA \ was obtained.
If we subtract the atomic radius of copper of $\frac{3.63}{2\sqrt 2}$\AA,
then we obtain a radius for Cl of 1.12 \AA. For Cl on Ag, we obtain
a radius of $2.62$\AA$-\frac{4.10}{2\sqrt{2}}$\AA = 1.17 \AA. This radius for
Cl is intermediate between the atomic (0.99 \AA) and ionic (Cl$^-$; 1.81 \AA))
\cite{Kittel}. The bond lengths predicted here and previously computed for 
the Cu(111)-$(\sqrt{3} \times \sqrt{3})\ {\rm R}30^\circ$-Cl
give a consistent picture of the radius of the adsorbed Cl$^-$ ion.

The Mulliken population analysis gave a slight charge transfer of 0.2 e to
0.25 $|e|$
to the chlorine atom, increasing in the order $fcc \sim hcp < bridge < atop$
; of course, one has to keep in mind that
the error associated with this number is 
larger in the case of metals with diffuse basis functions because the
diffuse Cl basis functions can be used to partially describe Ag states
and vice versa. Still, it is very interesting to compare the charge transfer
with the position of the Cl $3s$ eigenvalue (table \ref{Clpopulationtable}).
If we assume that intra-atomic interactions dominate the core eigenvalues
of the Cl atom, then the position of the Cl $3s$ eigenvalue
provides a measure of the atomic charge: the more negatively charged the Cl
ion, the lower the binding energy of the core eigenvalue. 
As this is related to the observable core
level shift, it can provide an experimental mechanism for distinguishing
adsorption sites.
We observe this correlation in our results for different adsorption sites
(table \ref{Clpopulationtable}), but
we find that this correlation is not retained when comparing Cl adsorption on Cu to that on Ag. The $3s$ eigenvalue is at lower binding energy for Cl on the
Ag surface
despite the Cl charge being lower in magnitude
 than that of Cl on the Cu surface.
This could be due to errors in the 
Mulliken charge estimates which are more reliable within a consistent
set of basis sets.

The position of the Cl levels 
is visible in the Mulliken projected densities of states (figures 
\ref{ClonAgdensity},\ref{DOSonClpxpypzbridge} and \ref{AgCuprojectedDOS}).
First, we projected the density of states on the chlorine orbitals
(figure \ref{ClonAgdensity}). 
The $3s$ level moves as described to higher energies 
 (corresponding to lower binding energies)
with
higher Mulliken orbital populations. 
For all adsorption sites, the valence band 
contribution consists of a broad
background from $\sim -0.2 E_h$ to $\sim -0.08 E_h$ which contains
contributions from Cl $3p_x,p_y$ and $3p_z$. The peak at $\sim -0.1 E_h$
is due to  $3p_x$ and $3p_y$ contributions in all cases; there is no distinct
peak from $3p_z$ contributions. The $3p_x$ peak and $3p_y$ peak are
virtually
identical in the case of fcc, hcp or atop adsorption sites; for the bridge site
the $p_y$ peak carries more weight than the $p_x$ peak 
(figure \ref{DOSonClpxpypzbridge}). 
These findings
are consistent with orbital projected Mulliken charges 
(table \ref{Clpxpypztable})
which are highest for $p_x$ and 
$p_y$ in the case of fcc, hcp and atop site. For the bridge site, the 
$p_y$  population is higher than $p_x$ and again higher than $p_z$
--- note that, in our choice of geometry, for the
bridge site, the Cl $3p_x$ orbital overlaps more with the silver atoms
in the top layer than the Cl $3p_y$ orbital 
(see figure \ref{fort34-figure}).
We can therefore conclude that in all sites    
the charge transferred to the Cl atom occupies those
 orbitals which have lowest overlap with the
silver atoms. This is consistent with an ionic rather than covalent picture
of the bonding.
We note that these orbitals may be separately observed by exploiting
the polarization dependence of angle resolved photoemission \cite{Lindsay}.
When comparing the chlorine adsorption on silver with
the chlorine adsorption on copper (figure \ref{AgCuprojectedDOS}), 
we find that the bandwidth of the 3$p$ states is similar;
as mentioned earlier the $3s$ peak is at
slightly higher energy
for Cl on Ag than for Cl on Cu, although the 
charge
is a bit smaller in magnitude 
(-0.207 $|e|$ for copper vs -0.198 $|e|$ for silver). The peak structure is
different from Cl on Ag(111) as we find one additional peak for Cl $3p_z$;
the two other peaks carry equal weight from Cl 3$p_x$ and Cl 3$p_y$.
As in the case of
silver, the Cl $3p_x$ and $3p_y$ orbitals carry slightly more charge than
Cl $3p_z$.

Finally, in figures \ref{Chargedensitytotal} and 
\ref{Chargedensitydifference}, we display the total charge density of Cl 
adsorbed in the fcc hollow site and the charge density difference. 
The plane was  chosen in such a way that the center of the chlorine
atom, the center of one of the nearest neighbor silver atoms (at a distance
of 2.62 \AA) and the centers of two more silver atoms (at a distance of
3.91 \AA) lie in
the same plane. The charge density difference has been obtained by
subtracting the charge densities of a clear Ag slab (with the geometry 
of the system Ag(111)-$(\sqrt{3} \times \sqrt{3})\ {\rm R}30^\circ$-Cl )
 and a layer of chlorine atoms from the charge density of the system after
the chlorine adsorption. We find no evidence of bond formation in the
charge density plots so that these plots support the idea of  an ionic 
binding mechanism. A further evidence of an ionic bonding is the
very small overlap population of $\sim$ 0.01 $|e|$ between Cl and the nearest
silver atom.

\section{Summary}

We have studied the adsorption of chlorine on the Ag(111) surface
using density functional theory calculations in which we have paid close
attention to numerical accuracy. The
fcc hollow site was found to be the preferred adsorption site, slightly
lower in energy than the hcp site. The adsorption energy is computed
to be 0.111 $E_h$ per atom with respect to free Cl atoms or 
0.123 $E_h$ per $\rm Cl_2$ molecule.

 The computed Ag-Cl distance of 2.62 \AA \ at the GGA-PW 
level (or 2.55 \AA \ at
the LDA level) are
intermediate between
two controversial values deduced from experiment. 
The 
 bond length is consistent with that computed for Cl on Cu which would 
indicate an effective radius of the Cl adsorbate of $\sim$ 1.15 \AA.
For the different adsorption sites
of Cl on Ag(111), we find that the charge on chlorine is
consistent with the relative position of the Cl $3s$ level, i.e. the higher
the charge on Cl, the lower the binding energy of the $3s$ level. This
might be valuable in the interpretation of core level shift observations.
From projected densities of states, we conclude that the chlorine charge
is larger in those orbitals which are most separated from silver which
means that the character of the bond tends towards ionic rather 
than covalent. 
In our choice of geometry, the chlorine $3p_x$ and $3p_y$ 
orbitals have a slightly higher occupation than the chlorine $3p_z$ orbital
for the threefold hollow sites
which might also be observable in photoemission experiments.

\section{Acknowledgments}
The authors would like to acknowledge support from EPSRC grant
GR/K90661.

\onecolumn

\newpage
\begin{table}
\begin{center}
\caption{\label{Agbulktable}
The ground state properties of bulk Ag. }
\vspace{5mm}
\begin{tabular}{ccccccc}
 & &  &  & \\
 & $a \ [{\rm \AA}]$ & $E_{coh} \ [E_h] $  & $B$ [GPa] \\
LDA  
 & 3.98 & 0.129 & 165 \\
GGA-PW, GGA-PBE
 & 4.10 & 0.088 & 113 \\
\\
LDA \cite{Asato} & 3.95 ... 4.01 & - & 137 ... 163  \\
LDA \cite{Khein} & 4.00 & - & 139 \\
GGA-PW \cite{Asato} & 4.12 ... 4.14 & - & 93 ... 98 \\
GGA-PW \cite{Khein} & 4.17 & - & 84.8 \\
experiment
 & 4.03 \cite{Khein} & 0.109 \cite{Gschneider} & 102 \cite{Khein} \\
\end{tabular}
\end{center}
\end{table}

\newpage
\begin{table}
\begin{center}
\caption{\label{Agsurfsummtable}The surface energy ($\frac{E_h}
{\rm surface \ atom}$) of
the low index silver surfaces.}
\begin{tabular}{ccccccc}
surface & LDA & GGA-PW, GGA-PBE  & Ref. \onlinecite{Vitosetal} (GGA-PBE) \\
(100) & 0.033 & 0.023 & 0.024 \\
(110) & 0.050 & 0.033 & 0.035 \\
(111) & 0.026 & 0.017 & 0.020 \\
\end{tabular}
\end{center}
\end{table}

\newpage
\begin{table}
\begin{center}
\caption{\label{ClonAgtable} Adsorption of Cl on the Ag(111) surface,
GGA-PW results. 
$\delta_{1-2}$
is the change in interlayer spacing between first and second silver layers
 relative to the bulk value, 
$d_{\rm {Cl-Ag \mbox{ }top\mbox{ } layer}}$ is the interlayer
 distance between the Cl  layer and the 
top Ag layer, $d_{\rm {Cl-Ag \ nn}}$ is the distance between
Cl and nearest neighbor Ag. The adsorption energy is the difference 
$E_{\rm {Cl \mbox{ }at \mbox{ } Ag(111)}}-{E_{\rm Ag(111)}-E_{\rm Cl}}$. 
$n$ is the number of silver layers in the slab.}
\begin{tabular}{ccccc}
$n$ &  $d_{\rm {Cl-Ag \mbox{ }top\mbox{ } layer}}$, in \AA & $\delta_{1-2}$,
in \AA &
$d_{\rm {Cl-Ag \ nn}}$, in \AA & $E_{adsorption}$ in $E_h$, per Cl atom \\
fcc site\\
3 & 2.01  & -0.043   & 2.62 & -0.11164 \\
4 & 2.01  & -0.039  &  2.62 & -0.11173 \\
hcp site\\
3 & 2.01 & -0.042 & 2.62 & -0.11068 \\ 
4 & 2.02 & -0.036 & 2.62 & -0.11129 \\
bridge site\\
 3 & 2.08 & -0.040 & 2.54 & -0.10878 \\
atop site\\
 3 & 2.38 & -0.034 & 2.38 & -0.09460 \\
\end{tabular}
\end{center}
\end{table}

\newpage
\begin{table}
\begin{center}
\caption{\label{ClonAgtableLDA}  Adsorption of Cl on the Ag(111) surface,
LDA results. The notation is the same as in table \ref{ClonAgtable}}
\begin{tabular}{ccccc}
site & $d_{\rm {Cl-Ag \mbox{ }top\mbox{ } layer}}$, in \AA & $\delta_{1-2}$,
in \AA &
$d_{\rm {Cl-Ag \ nn}}$, in \AA & $E_{adsorption}$ in $E_h$, per Cl atom \\
fcc & 1.96 & -0.052 & 2.55 & -0.13653 \\
hcp & 1.96 & -0.046 & 2.55 & -0.13422 \\
bridge & 2.03 & -0.052 & 2.47 & -0.13297 \\
atop & 2.33 & -0.047 & 2.33 & -0.11487 \\
\end{tabular}
\end{center}
\end{table}

\newpage
\begin{table}
\begin{center}
\caption{Charge and position of $3s$ eigenvalue
for Cl on different adsorption sites.}
\label{Clpopulationtable}
\begin{tabular}{ccc}
site & charge, in $|e|$ & $3s$ level relative to $E_{\rm Fermi}$, in $E_h$ \\
\multicolumn{3}{c}{Cl on Ag:}\\
fcc    & -0.198 & -0.563 \\
hcp    & -0.204 & -0.562 \\
bridge & -0.218 & -0.555 \\
atop    & -0.252 & -0.532 \\
\multicolumn{3}{c}{Cl on Cu:}\\
fcc    & -0.207 & -0.578 \\
\end{tabular}
\end{center}
\end{table}

\newpage
\begin{table}
\begin{center}
\caption{Orbital projected Cl charge, in $|e|$, on different adsorption sites.}
\label{Clpxpypztable}
\begin{tabular}{cccc}
site & $p_x$  & $p_y$  &  $p_z$  \\
\multicolumn{3}{c}{Cl on Ag:}\\
fcc    & -3.794 & -3.794 & -3.629 \\
hcp    & -3.795 & -3.795 & -3.636\\
bridge & -3.766 & -3.840 & -3.628 \\
atop    & -3.822 & -3.822 & -3.605 \\
\multicolumn{3}{c}{Cl on Cu:}\\
fcc    & -3.802 & -3.802 & -3.598\\
\end{tabular}
\end{center}
\end{table}

\newpage
\begin{figure}
\caption{Projected density of states for Cl on Ag(111) and different
adsorption sites.
The part of the density of states which is due to the Cl basis functions,
is shown in this projection. The Fermi energy is fixed at 0 $E_h$.}
\label{ClonAgdensity} 
\end{figure}

\centerline{\psfig{figure=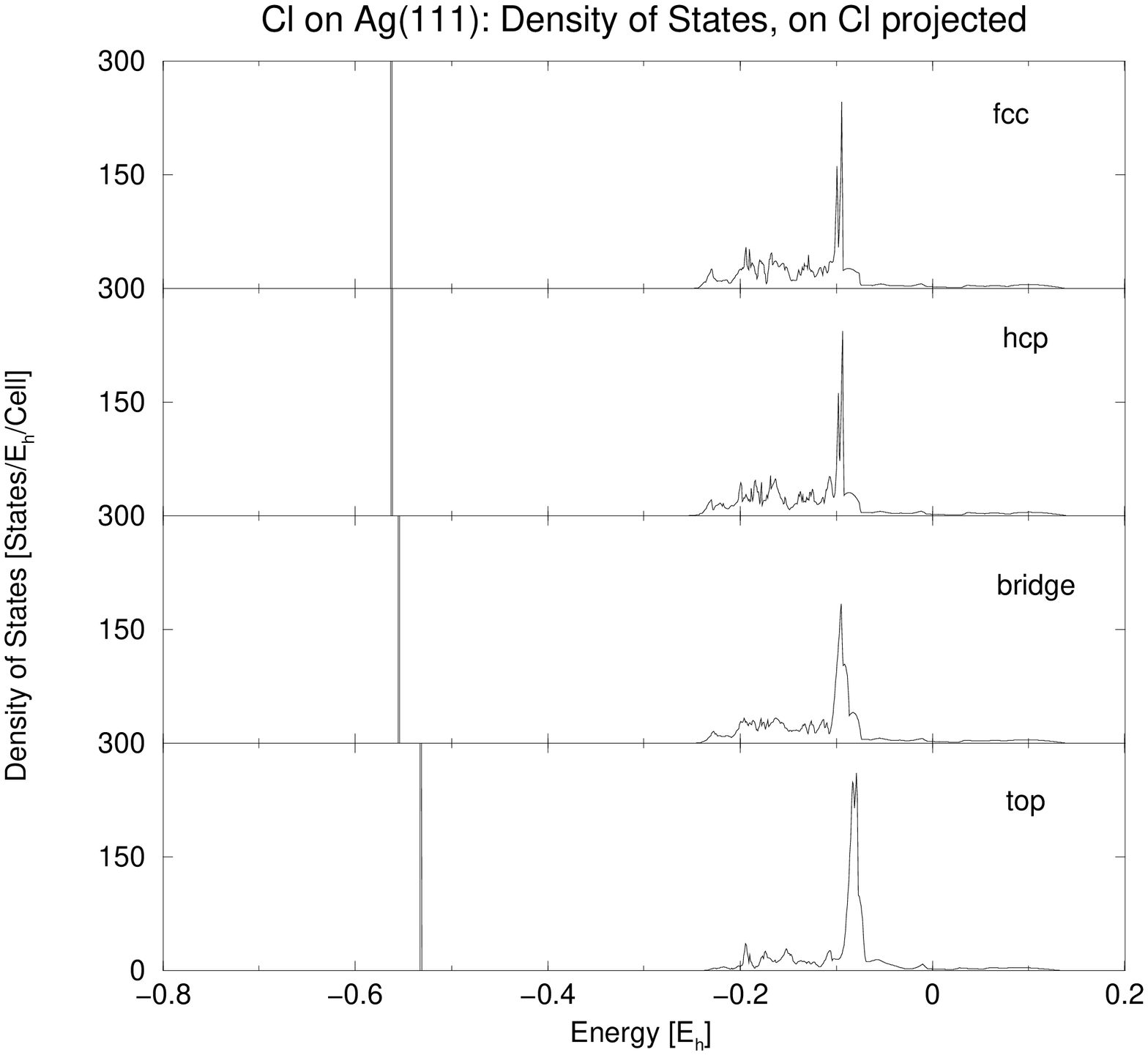,width=15cm,angle=270}}
\vfill

\newpage
\begin{figure}
\caption{Projected density of states for Cl on Ag(111), bridge site.
The density of states is shown, 
projected on the Cl basis functions ($p_x, p_y$ and $p_z$). 
The Fermi energy is fixed at 0 $E_h$.}
\label{DOSonClpxpypzbridge} 
\end{figure}

\centerline{\psfig{figure=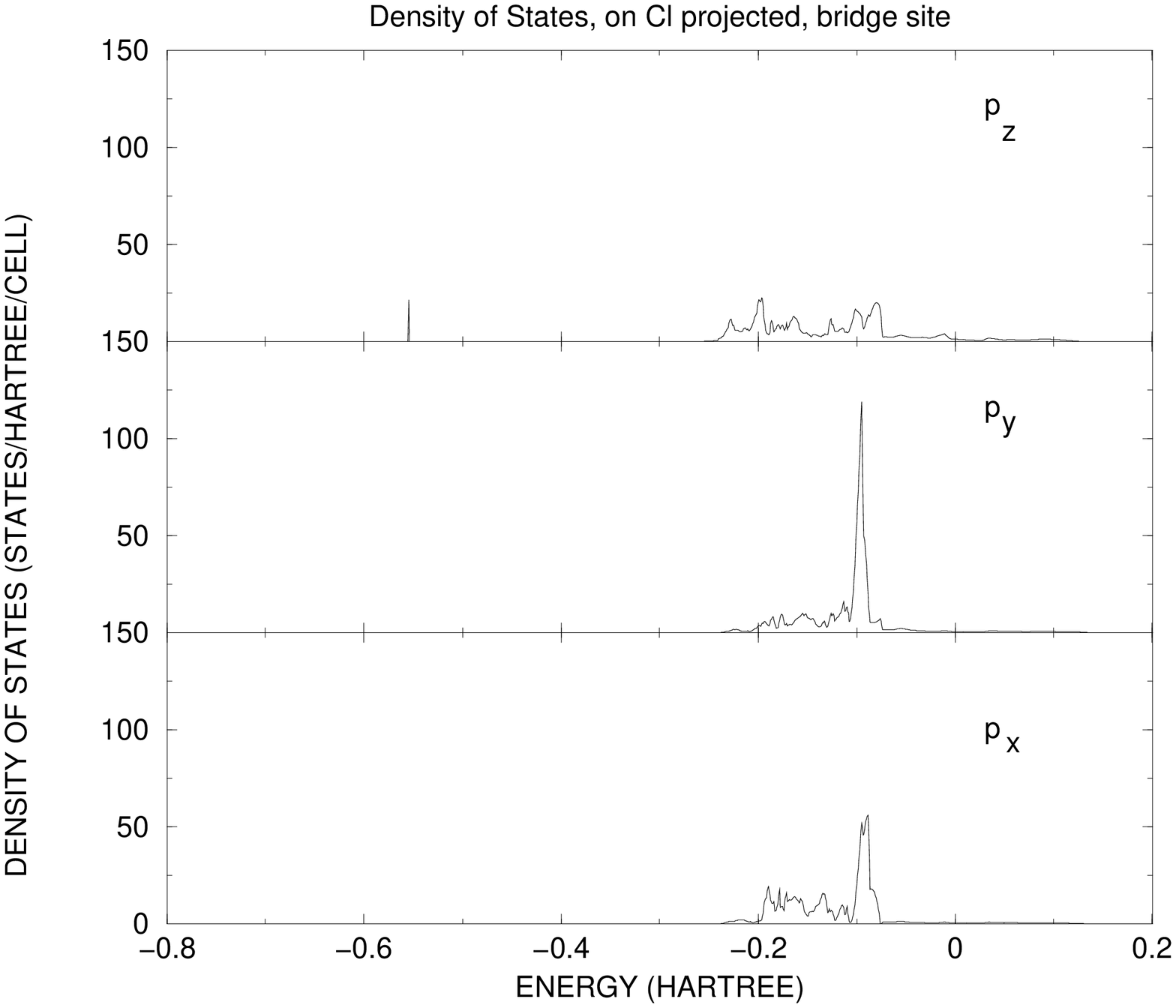,width=15cm,angle=270}}
\vfill

\newpage
\begin{figure}
\caption{Projected density of states for Cl on Cu(111) and Cl on Ag(111),
fcc hollow.
The part of the density of states which is due to the Cl basis functions,
is shown in this projection. The Fermi energy is fixed at 0 $E_h$.}
\label{AgCuprojectedDOS} 
\end{figure}

\centerline{\psfig{figure=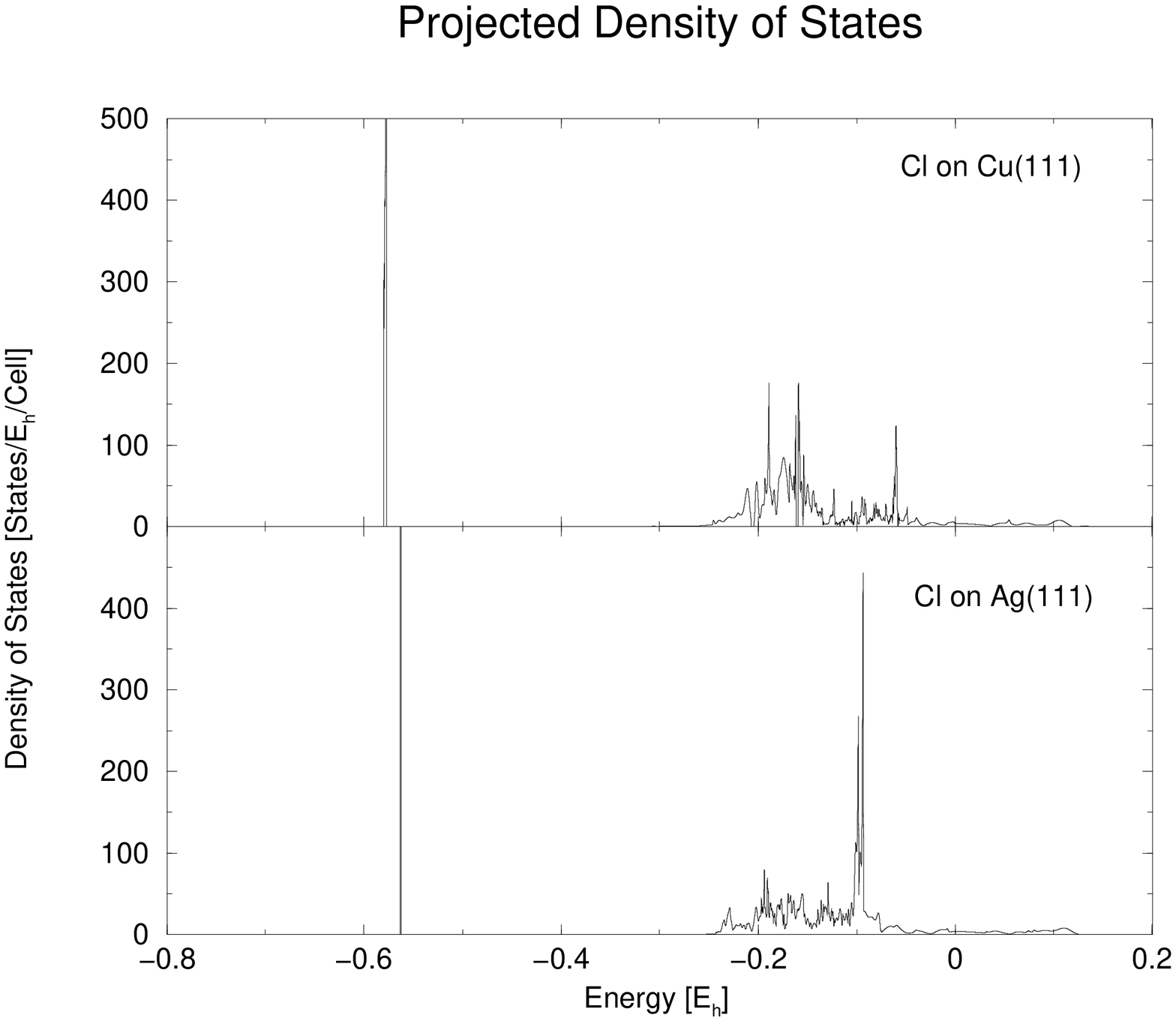,width=15cm,angle=270}}
\vfill

\newpage
\begin{figure}
\caption{Total charge density for Cl on Ag(111), fcc hollow, from 0.01 to
0.09 a.u. in steps of 0.01.}
\label{Chargedensitytotal} 
\end{figure}

\centerline{\psfig{figure=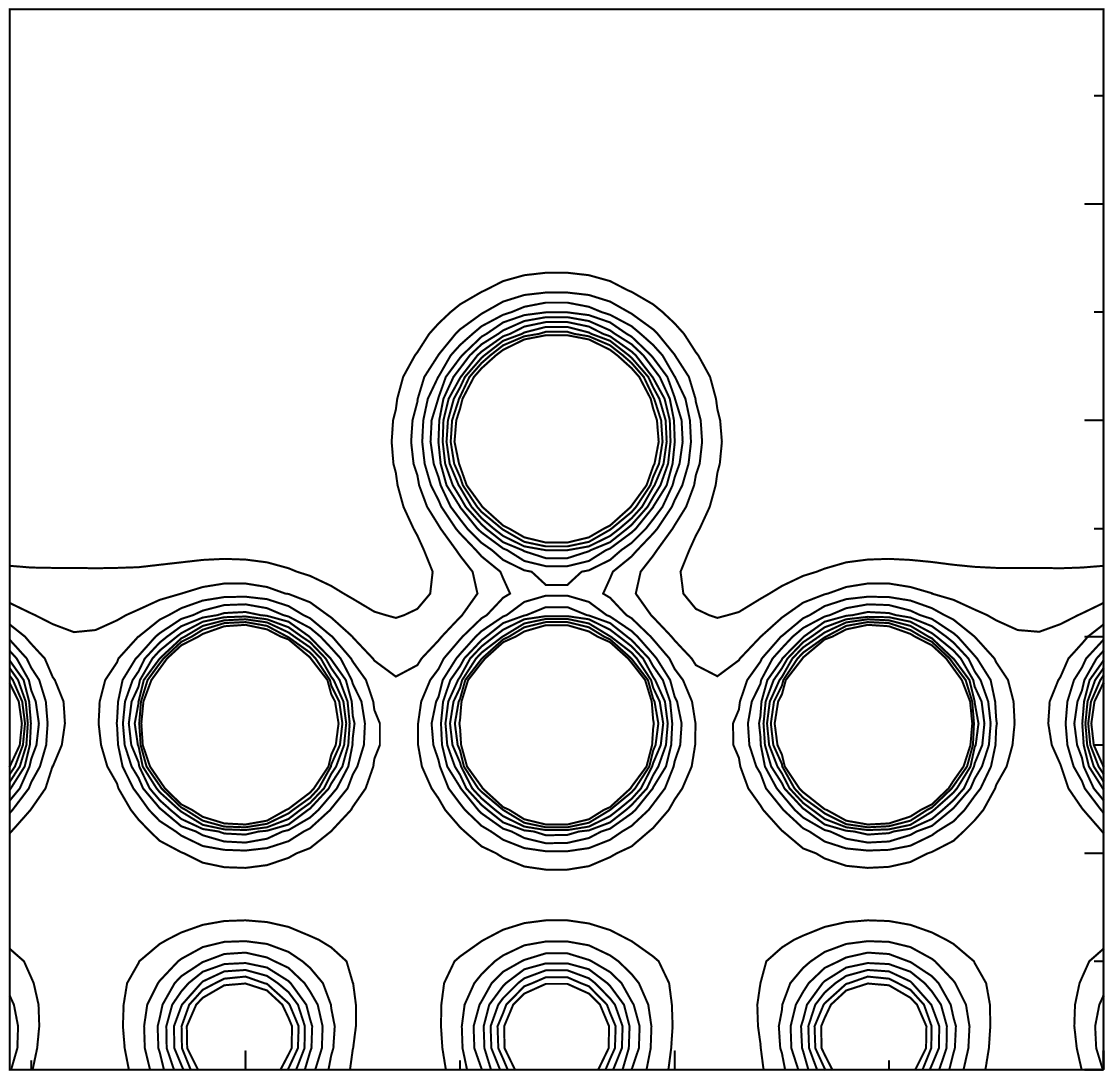,width=15cm,angle=0}}
\vfill

\newpage
\begin{figure}
\caption{Difference in charge density 
(i.e. charge Cl on Ag(111) - charge of clear Ag(111) - charge of a layer of
 Cl atoms)
for Cl on Ag(111), 
fcc hollow, from -0.007 to
0.009  in steps of 0.001 a.u. 
Excess negative charge 
is displayed with long dashed
lines, excess positive charge with solid lines, zero difference with short
dashed lines}
\label{Chargedensitydifference} 
\end{figure}

\centerline{\psfig{figure=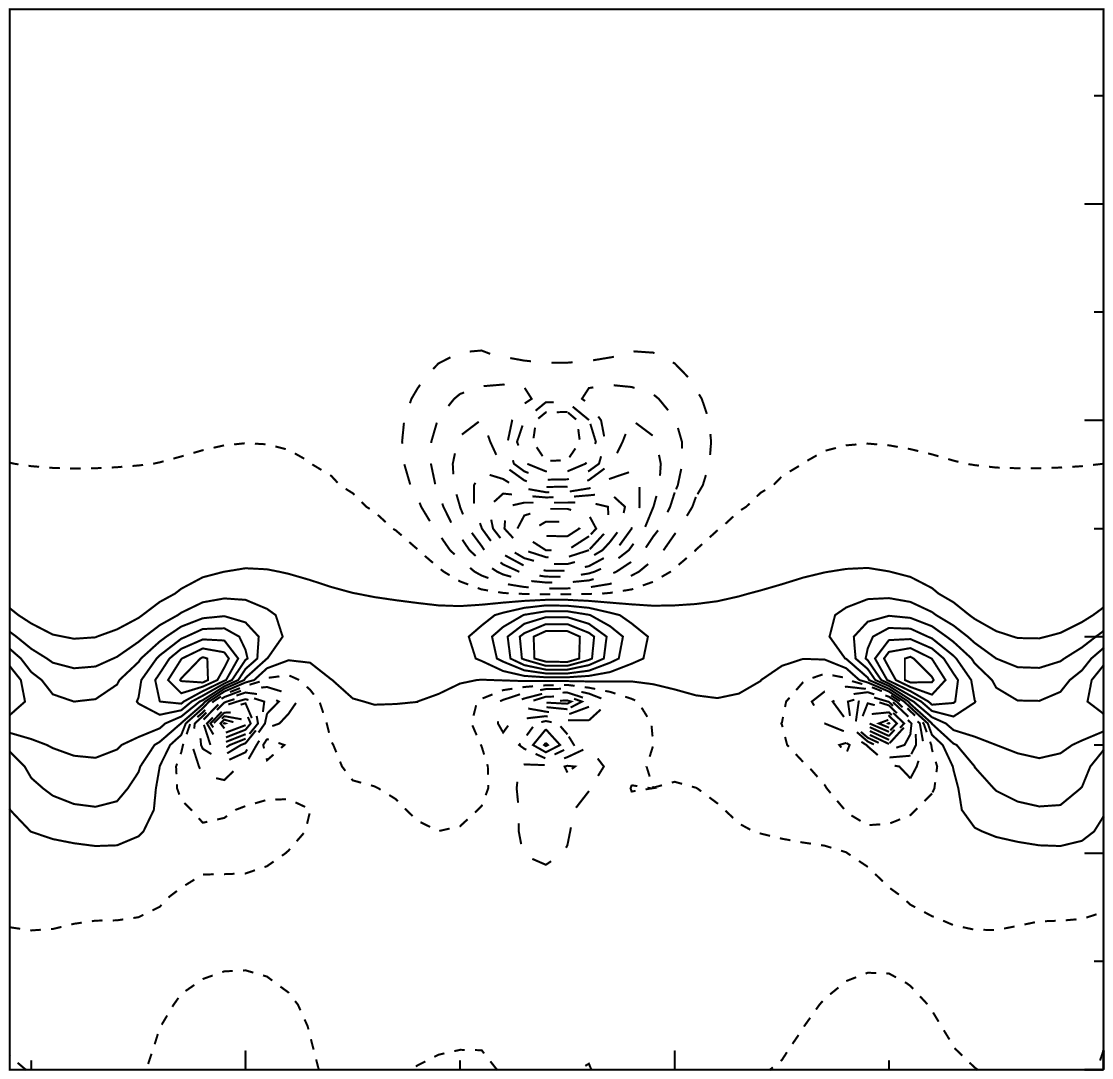,width=15cm,angle=0}}
\vfill
\end{document}